# High Responsivity Gate Tunable UV-Visible Broadband Phototransistor Based on Graphene –WS$_2$ Mixed Dimensional (2D-0D) Heterostructure

Shubhrasish Mukherjee[1], Didhiti Bhattacharya[1], Sumanti Patra[1], Sanjukta Paul[1], Rajib Kumar Mitra[1], Priya Mahadevan[1], Atindra Nath Pal*[1] and Samit Kumar Ray*[1,2]

[1] *S. N. Bose National Center for Basic Science, Sector III, Block JD, Salt Lake, Kolkata – 700106*

[2] *Indian Institute of Technology Kharagpur, 721302, West Bengal, India*

**Abstract:**

Recent progress in the synthesis of highly stable, eco-friendly, cost-effective transition metal-dichalcogenides (TMDC) quantum dots (QDs) with their broadband absorption spectrum and wavelength selectivity features have led to their increasing use in broadband photodetectors. With the solution based processing, we demonstrate a super large (~0.75 mm$^2$), UV-Vis broadband (365-633 nm), phototransistor made of WS$_2$ QDs decorated CVD graphene as active channel with extraordinary stability and durability in ambient condition (without any degradation of photocurrent till 4 months after fabrication). Here, colloidal 0D WS$_2$-QDs are used as the photo absorbing material and graphene acts as the conducting channel. A high photoresponsivity (3.1 × 10$^2$ A/W), higher detectivity (2.2×10$^{12}$ Jones) and low noise equivalent power (4×10$^{-14}$ W/Hz$^{0.5}$) are obtained at a low bias voltage ($V_{ds}$ = 1V) at an illumination of 365 nm with an optical power as low as 0.8 µW/cm$^2$, which can further be tuned by modulating the gate bias. While comparing the photocurrent between two different morphologies of WS$_2$ (QDs and 2D nanosheets), a significant enhancement of photocurrent is observed in case of QDs based device. *Ab initio* density functional theory based calculations further support our observation, revealing the role of quantum confinement for the enhanced photo response. Our work reveals a strategy towards making a scalable, cost-effective, highly performing hybrid two-dimensional (2D/0D) photo detector with graphene-WS$_2$ QDs, paving the way towards the next generation optoelectronic applications.

**Introduction**

Photodetectors that can convert light into electrical signals are at the heart of technologies that affect our daily lives[1,2,3]. Sensitive detection of light of different wavelengths has various applications including optical communications, video-imaging, security, night-vision, optical data storage, biomedical imaging etc. In traditional photodetectors, various photosensitive semiconductors are extensively used, such as ZnO[4], GaN for UV (< 400 nm)[5], Si for visible-NIR (~450-1100 nm)[6], Ge for NIR–mid-IR (~1-5 µm)[7], PbS for IR (~700-1400 nm)[8], HgCdTe for MIR to far-IR (>5 µm)[9] etc. However, such materials have disadvantages like high production cost, environmental toxicity, complicated microfabrication and high thermal budget processes, cryogenic operating temperature etc. Two dimensional (2D) layered materials (graphene[10], TMDCs[11] etc.) offer a new viable alternative due to their promising electronic and optical properties. While graphene is considered as an outstanding channel material for a transistor due to its ultrahigh charge carrier mobility (~up to 60000 cm$^2$V$^{-1}$S$^{-1}$, at room temperature on a substrate)[12], it has limitations in the field of optoelectronics because of its gapless nature, low absorption cross-section and short photogenerated carrier lifetime[13,14]. There have been attempts to increase the light absorption in graphene using plasmonic coupling[15], or microcavity confinement[16]. However, the most successful approach is by creating heterostructures and grafting a photo absorbing material on top of the graphene devices. For example, the heterostructures of graphene with PbS[17] or other quantum dots (ZnO[18], Si[19] etc.) have shown improved photoresponsivity. Also, the formation of vertical heterostructure (2D-2D) of graphene and layered transition metal dichalcogenides (TMDCs)[20,21] such as $MoS_2$, $WS_2$ can lead to very high quantum efficiency upon light illumination due to effective photoexcited carrier separation at the interface. However, the fabrication of these devices is expensive and needs delicately controlled sample transfer technique, which has low-yield and multiple lithography procedures.

Graphene-based photodetectors coupled with colloidal quantum dots (QDs)[22,23,24] ( PbS, CdS, Perovskite etc.) have emerged as a feasible alternative due to their good detectivity, fast response time, high gain, and low production cost. However, most of the synthesis processes of QDs are hazardous, complex in reaction kinetics and most importantly suffer from stability issues[25]. To circumvent these difficulties, solution processed TMDC QDs[26,27] are found to be suitable for fabrication of graphene based broadband, highly stable photodetectors because of their broad absorbance band, direct and tunable band gap, easy synthesis process and higher stability. Among TMDCs, $WS_2$ is an emerging candidate for its outstanding optoelectronic properties. The development of a facile and simple synthesis procedure for the preparation of the crystalline $WS_2$ QDs[28] has opened up the possibilities to create graphene-$WS_2$ QDs based phototransistor, which is still unexplored for its optoelectronic applications.

In this work, we demonstrate the largest area (~0.75 $mm^2$) (to the best of our knowledge) lithography free, cost effective hybrid phototransistor based on CVD graphene and chemically exfoliated $WS_2$ quantum dots. In this 2D-0D configuration, $WS_2$ QDs act as light absorber, while the CVD grown single layered graphene is the conductive channel for current flow. This device, having a large photoactive area (~0.75 $mm^2$) provides UV-visible (~365-633 nm) broadband photoresponse with good gate tunability and extra-ordinary stability. The fabricated phototransistor exhibits very high responsivity ($3.1 \times 10^2$ A/W), high detectivity ($2.2 \times 10^{12}$ Jones) and low noise equivalent power ($4 \times 10^{-14}$ W/$Hz^{0.5}$) at a low bias voltage ($V_{ds}$ = 1V and $V_{bg}$ = 0V) at an illumination of 365 nm with an optical power of 0.8 μW/$cm^2$. We further perform a comparative study of the photoresponse between two different morphologies of $WS_2$ ($WS_2$ quantum dots (0D) and 2D $WS_2$ nanosheets) in same device configurations. We explain our observation through first principle calculations and show that the increased localization of the wave function due to quantum confinement effects is responsible for the enhancement of optical attributes in quantum dots. The $WS_2$ QDs with graphene offers a higher photoresponse than its 2D ($WS_2$ nanosheets) counterpart demonstrating a highly stable, low cost graphene based UV-visible (365-633 nm) broadband phototransistor. Adding with the simple scalable solution process for the preparation of $WS_2$ QDs, these results are very promising for wafer-scale photodetector devices for future optoelectronic applications.

## Results and discussions

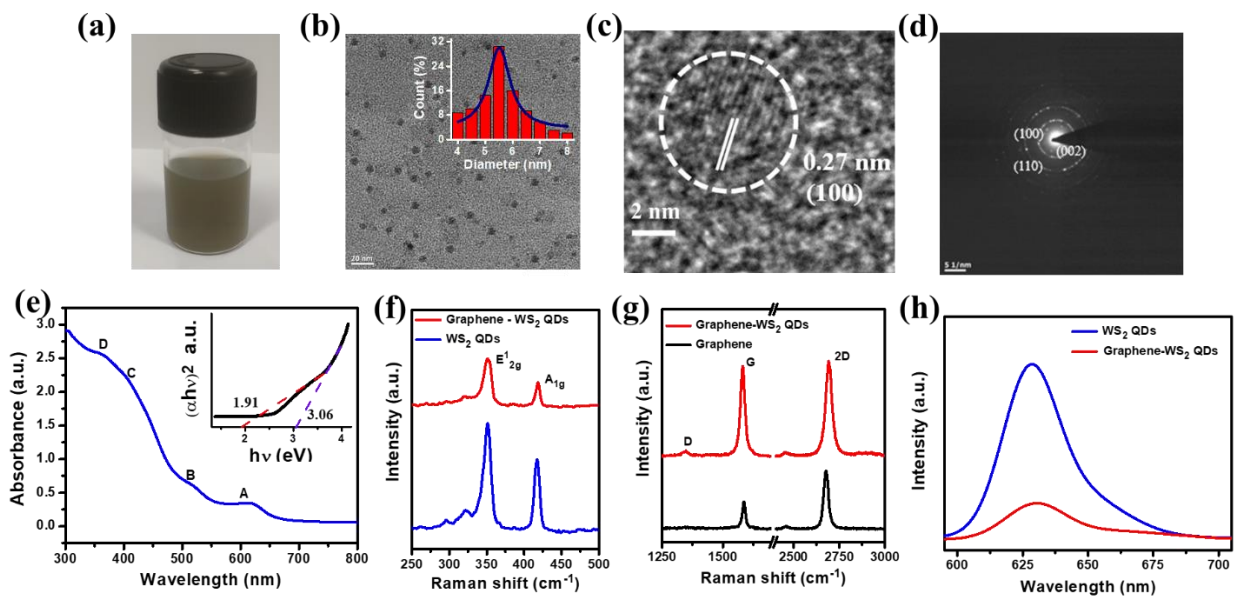

**Figure 1:** Characterizations of $WS_2$ QDs and graphene-$WS_2$ QDs heterostructures **(a)** Optical image of synthesized $WS_2$ QDs in IPA solution **(b)** TEM image of the $WS_2$ QDs with size distribution of the QDs (Inset) showing the average radius of $WS_2$ QDs to be ~ 2.75 nm **(c)** HRTEM image of $WS_2$ QDs showing the lattice fringes with an interlayer spacing 0.27 nm **(d)** The selected area electron diffraction pattern of $WS_2$ QDs **(e)** UV-visible absorbance spectra of the $WS_2$ QDs and the Tauc plot based on the absorbance data (Inset) **(f)** Raman spectra of pristine $WS_2$ QDs and the fabricated hybrid $WS_2$ QDs/graphene heterostructure **(g)** Raman spectra of graphene before and after making the hybrid heterostructure **(h)** Photoluminescence spectra of pristine $WS_2$ QDs and the heterostructure.

$WS_2$ QDs were synthesized by sono-chemical exfoliation using Li-intercalation technique[28]. This low cost, eco-friendly exfoliation technique without any inert environment, using stabilizing reagents is able to produce colloidal semiconducting (2H-phase) $WS_2$ QDs which can easily be dispersed in appropriate solvents[29]. The optical image of as-synthesized $WS_2$ QDs dispersed in isopropanol (IPA) solution is shown in **Figure 1a** and to investigate the morphology of the as synthesized $WS_2$ QDs, high-resolution transmission electron microscopy (HRTEM) analysis has been carried out (**Figure 1b, c**). The micrograph of the sample with an average QDs size (radius) ~2.75 nm is shown in **Figure 1b** and the inset figure indicates the size distribution of the $WS_2$ QDs. From the lattice fringes in HRTEM image (**Figure 1c**), the interlayer spacing is found to be 0.27 nm which reveals the formation of (100) plane in semiconducting 2H-$WS_2$[28,30]. Selected area electron diffraction (SAED) pattern shown in **Figure 1d** indicates the crystallinity of the synthesized $WS_2$ QDs. The X-ray diffraction (XRD) pattern (**Figure S1a**) also supports the formation of crystalline 2H-phase of $WS_2$ QDs[28,31]. From a typical UV-visible absorption spectrum (**Figure 1e**) of $WS_2$ QDs (300-800 nm), the characteristic exciton peaks are observed at ~ 616 nm (A), ~516 nm (B), arising due to the transitions between the conduction band minima and spin-orbit-coupling induced splitting of the valence band levels[32,33] Also, the excitonic C (~ 407 nm) and D (~ 366 nm) peaks are present in this synthesized $WS_2$ QDs, which are attributed to the Van Hove singularities leading to strong absorption[34,35]. These QDs possess enhanced bandgap (compared to its bulk counterpart) due to the phenomena of quantum confinement and we estimate the band gap from the Tauc plot (inset of **Figure 1e)**, depicting it as a direct band gap semiconductor[36,37](emission at 1.97 eV for 532 nm excitation) having a broad absorption band (1.91-3.06 eV). The Raman spectra of $WS_2$ QDs are shown in **Figure 1f** (bottom panel). Two main characteristics peaks at 351.1 and 417.2 cm$^{-1}$ are observed, corresponding to the out-of-plane ($E^1_{2g}$) and in-plane modes of vibration ($A_{1g}$) , respectively[38,39]. The Raman features remain unaffected in the hybrid structure (**Figure 1f,** top panel), indicating the good crystallinity of $WS_2$ QDs even in the presence of graphene. **Figure 1g** shows the Raman spectra of bare graphene[40] (bottom panel) and graphene-$WS_2$ QDs (top panel) heterostructure. The absence of significant D peak at 1350 cm$^{-1}$ in graphene reveals that negligible defects are introduced in the solution grown heterostructure device. Additionally, a blue shift of the 2D peak position (from 2680 cm$^{-1}$ to 2694 cm$^{-1}$), along with a significant enhancement of both the G (~3.2) and 2D (~1.67) peak intensities are also observed in the heterostructure, indicating the injection of photo-generated holes from $WS_2$ QDs to graphene[41]. This enhancement of the peak intensity has also been observed in previously reported graphene-metal nanoparticles, semiconducting quantum dot system and indicates the possibility of localized surface plasmon resonance (LSPR) effect[42,43]. From the photoluminescence (PL) spectra (excitation 532 nm) of $WS_2$ QDs, a strong emission peak is observed (**Figure 1h)** at ~628 nm which corresponds to a direct band gap of 1.97 eV[44,32]. It is observed that the emission intensity of graphene-$WS_2$ QDs heterostructure is quenched significantly as compared to intrinsic $WS_2$ QDs which also corroborates the charge transfer process across the heterostructure interface[45,46].

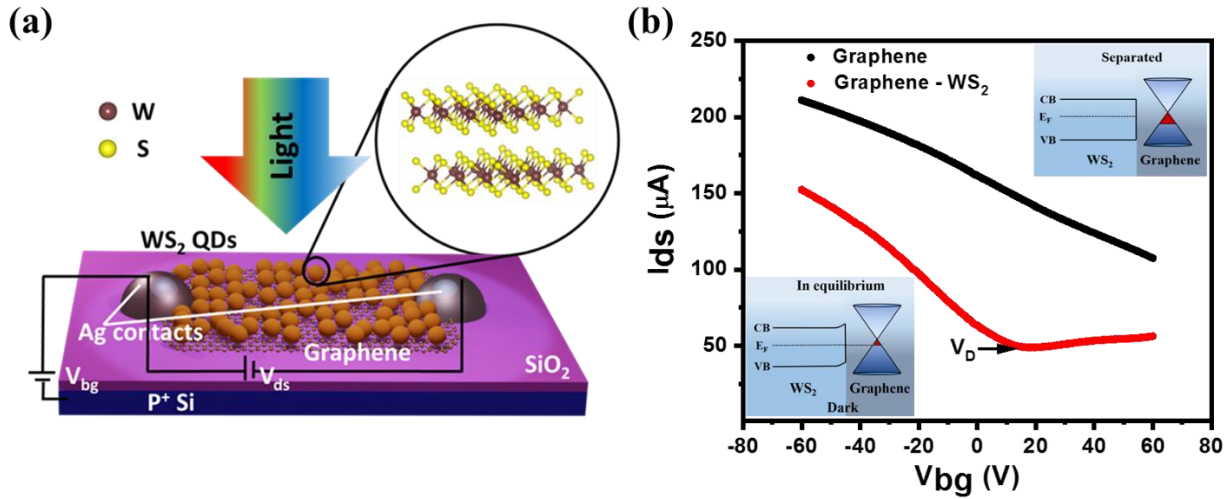

**Figure 2: (a)** Schematic of the back-gated graphene-$WS_2$ QDs hybrid device **(b)** Transfer characteristics of the graphene transistor before and after deposition of $WS_2$ QDs. The energy band diagrams before and after the formation of heterostructure are shown in the insets.

The schematic of a hybrid graphene–$WS_2$ quantum dots phototransistor is shown in **Figure 2a (SEM image Figure S2a),** where large area CVD graphene acts as the carrier transport channel, and $WS_2$ QDs are used as the photon absorbing material. Gate voltages are applied to p+-doped Si with 300 nm thick $SiO_2$ acting as a gate dielectric. **Figure 2b** compares the transfer characteristics of the field effect device (in dark) before and after the deposition of quantum dots. In the pristine device, no charge neutrality point (CNP) is observed, possibly due to the hole doping due to the PMMA based wet transfer and consequent adsorption of water molecules by the graphene film [47]. The deposition of $WS_2$ QDs does not alter the nature of the transfer characteristics, indicating that graphene is the dominant carrier transport channel, but a dramatic shift of the CNP ( > 60V to 18V) is observed, confirming the significant electron transfer from $WS_2$ quantum dots. The hole mobility of the hybrid device is found to be ~ 400 $cm^2/VS$ (**Figure S2b**)[48]. The transfer of electrons from $WS_2$ QDs to graphene forms a built-in electric field, leading to band bending in $WS_2$ to equilibrate the Fermi levels[17,19], as shown schematically by the energy band diagrams in the inset of **Figure 2b**.

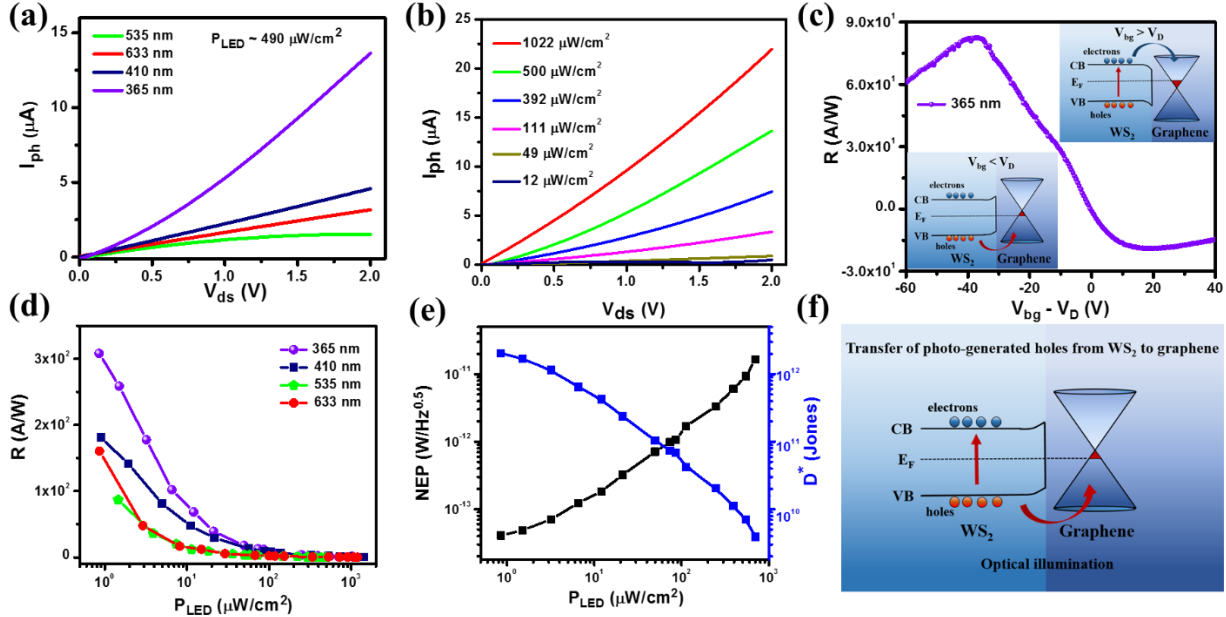

**Figure 3:** **(a)** Photocurrent ($I_{ph}$) as a function of $V_{ds}$ at $V_{bg}$ = 0V under UV-visible illuminations at $P_{LED}$ ~ 470 $\mu W/cm^2$ **(b)** $I_{ph}$ as a function of $V_{ds}$ at $V_{bg}$ = 0V under different LED powers at λ = 365 nm **(c)** Responsivity (R) as a function of $V_{bg}$-$V_D$ under illumination of ~ 3.21 $\mu W/cm^2$ at λ = 365 nm, $V_{ds}$ = 0.1 V. The band bending of the graphene-$WS_2$ QDs at $V_{bg}$ - $V_D$ < 0 and $V_{bg}$ - $V_D$ > 0. (Insets) **(d)** Responsivity (R) as a function of optical powers at different wavelengths at $V_{ds}$ = 1V, $V_{bg}$ = 0V **(e)** Noise equivalent power (NEP) and the Specific detectivity ($D^*$) of the device as a function of optical powers at λ = 365 nm **(f)** Energy band diagram showing the electron–hole generation in graphene-$WS_2$ QDs interface and charge transfer under the UV-visible illumination.

**Figure 3a** shows the photocurrent ($I_{ph}$) as a function of the drain-source bias ($V_{ds}$) for the hybrid device under the UV-visible (~365-633 nm) illumination of same optical power ($P_{LED}$ ~ 490 $\mu W/cm^2$), at a gate voltage ($V_{bg}$) of 0V at room temperature. The maximum photocurrent is obtained at λ=365 nm (at an optimum concentration of $WS_2$ QDs of ~ 2.25 mg/ml, **Figure S3b**). The power dependent photocurrent results are presented in **Figure 3b** at a particular wavelength (λ = 365 nm). A significant photocurrent can be detected with a power as low as ~0.8 $\mu W/cm^2$ and a monotonic increase of photocurrent is observed with the increase of incident power. Photo-responsivity (R) is one of the most important figure of merits to evaluate the performance of photodetectors which is a measure of a device's electrical response and is defined as[49]

$$R = \frac{I_{ph}}{P_{LED}} \quad \text{...........................} (1)$$

where, $I_{ph}$ is the photocurrent defined by ($I_{light}$ - $I_{dark}$) and $P_{LED}$ is the optical illumination power on the surface of the device. The transfer characteristics of the hybrid device both in dark and in presence of 365 nm illumination are depicted in **Figure S3a**. **Figure 3c** represents the responsivity (R) of our device as a function of the applied back-gate voltage ($V_{bg}$) (at λ = 365 nm and $V_{ds}$ = 0.1V). It is observed that the responsivity changes from positive to negative as the gate voltage sweeps through the Dirac point ($V_{bg}$ - $V_D$ = 0). For $V_{bg}$ < $V_D$, holes are responsible for the charge transport in graphene. With decrease of $V_{bg}$, the Fermi level of graphene is lowered and it enhances the built-in electric field at the interface of the heterostructure. This facilitates more photo-generated holes to be transferred from the QDs to graphene (inset, **Figure 3c**), resulting in an enhanced responsivity. Upon further decrease of $V_{bg}$, the responsivity shows a maximum at $V_{bg}$-$V_D$ ~ -40V and decreases further. The trapped electrons in $WS_2$ QDs start tunneling through the thinned barrier to graphene, leading to the decrease of the responsivity at higher negative gate voltages. Near the Dirac Point, the photoresponsivity becomes zero, that is, the device can effectively be turned "off" by tuning the gate voltage. At, $V_{bg}$ > $V_D$ the electrons are the dominant charge carrier and the responsivity becomes negative. The negative responsivity again increases with increasing $V_{bg}$ due to the enhanced injection of electrons from QDs to graphene (Inset **Figure 3c**). Similar to the hole side, a maxima in responsivity is observed at lower ($V_{bg}$ - $V_D$) value (~ 15V), possibly due to the lower mobility of electrons compared to the holes in the hybrid device. Thus, the gate voltage provides an additional functionality to switch the sign of the responsivity which demonstrates the potential of this device as a back-gate-tunable ultrahigh responsive phototransistor. This tunability is of great importance in photodetectors because it allows to control the state (on/off) of the detector as well as to adjust the required responsivity, depending on the light intensity to be detected.

**Figure 3d** shows the variation of responsivity with the illumination power at $V_{ds}$ = 1V and $V_{bg}$ = 0V for four different wavelengths. We observe an extremely high responsivity of ~3.1 × $10^2$ A/W for λ = 365 nm with $P_{LED}$ = 0.8 µW/cm². With increasing $P_{LED}$, R decreases significantly, typical characteristics of the photodetectors[20,50,51]. More importantly, a significantly high responsivity is also observed for different spot wavelengths of 410 nm, 535 nm and 633 nm, making the efficacy of our device as UV-visible broadband phototransistor. Responsivity is found to be ~ 1.81 × $10^2$ and 1.6 × $10^2$ A/W for 410 and 633 nm illumination, respectively for an optical power of 0.9µW/cm², while R is ~ 0.87 × $10^2$ A/W for 535 nm with a power of ~1 µW/cm². There are two main mechanisms of photocurrent either through the bound excitons or due to the photo-generated free charge carriers[52]. It is noted that the responsivity decreases from UV (365 nm) to green (535 nm) illumination by following the pattern of absorption spectrum of $WS_2$ QDs. It increases slightly at 633 nm due to the presence of strong exciton (A) peak, as observed in both absorbance and photoluminescence spectra. The responsivity of the fabricated graphene-$WS_2$ quantum dots hybrid device is significantly higher than that reported for pristine $WS_2$ (1.88 × $10^{-2}$ A/W)[53] and graphene (0.61 × $10^{-2}$ A/W)[54] based devices as well as some of the graphene based hybrid photodetectors[55,56,57,58,59] (see **Table 1**). These results illustrate that superior large area broadband photo-transistor can be achieved by using graphene and $WS_2$ quantum dots.

The figure of merits used to evaluate the capability of the detection of minimum optical illumination for a photodetector are the noise equivalent power (NEP) and the specific detectivity ($D^*$). The theoretical limit of the noise equivalent power (NEP) of these devices has been estimated by using the equation [60]

$$NEP = \frac{S_n}{R} \quad \text{...............................} \quad (2)$$

Where, $S_n$ is the total current noise and R is the photo-responsivity of the device.
The specific detectivity can be found out from the equation, [60]

$$D^* = R\frac{A^{\frac{1}{2}}}{S_n} \quad \text{................................} \quad (3)$$

Assuming the dark current to be dominated by the shot noise, the other sources of noises i.e. Johnson and flicker noises have been ignored here. Thus the spectral dependence of the specific detectivity then can be represented as[61]

$$D^* = R\left(\frac{A}{2eI_{dark}}\right)^{\frac{1}{2}}, \quad \text{.....................} \quad (4)$$

where, $I_{dark}$ is the dark current of the device at $V_{ds} = 1V$ and $V_{bg} = 0V$.
Using these equations, we get the limit of NEP for our devices to be $\sim 4 \times 10^{-14}$ W Hz$^{-1/2}$, which is presented in **Figure 3e** for $\lambda = 365$ nm. This low value of NEP indicates that the graphene-WS$_2$ QDs phototransistor may be very proficient for weak light detection. Since specific detectivity (D$^*$) is area-normalized, it is used as the yardstick for comparing the performance of various detectors. D$^*$ (also depicted in **Figure 3e**) is also moderately high ($\sim 2.2 \times 10^{12}$ Jones) ( NEP and D$^*$ for other wavelengths are presented **Figure S4**), comparable to some of the similar devices reported in the literature[17,62,63,21] (**Table 1**).

In order to understand the detailed operating mechanism of the hybrid graphene-WS$_2$ phototransistor, the schematic band alignment is presented in **Figure 3f**. Upon illumination, electron-hole pairs are generated in WS$_2$ QDs and separated at the interface by the induced electric field caused by the band bending due to the work-function mismatch between graphene and WS$_2$ QDs (as discussed previously). The photogenerated holes get easily transferred to the graphene layer due to the initial upward band bending, and the electrons are trapped in WS$_2$ QDs with a longer lifetime leading to photogating effect[17]. Consequently, positive charges (holes) in the graphene sheet are re-circulated for many times resulting in a high photoresponsivity in the hybrid device.

The temporal photocurrent characteristics in the broad UV-visible wavelength range (365-633 nm), with $V_{ds} = 1V$, $V_{bg} = 0V$ for a constant optical power ~470 µW/cm$^2$ are shown in **Figure 4a**. It is found that the photocurrent can be effectively switched "on" and "off" with very good repeatability. The increased photocurrent of graphene-WS$_2$ hybrid phototransistor for different wavelengths is in well agreement with the absorbance spectra of WS$_2$ QDs. Upon illumination of $\lambda = 365$ nm, this device exhibits a large positive photocurrent of about ~7 µA which can further be modulated by varying $V_{bg}$, as shown in **Figure 4b**. The temporal photoresponse of a photodetector is characterized by its response time. The response time, including rise and decay time ($\tau_1$ and $\tau_2$), of a photodetector is defined as the time interval for the current changes from 10% to 90% when light is turn on or off. The rise and decay times are determined to be 1.2 sec and 1.32 sec, respectively (**Figure 4c**), which are very much comparable with other graphene based hybrid photodetectors[64,65,55,66]. Furthermore, the stability of the device is checked by measuring the same device after 4 months. No significant degradation of photocurrent is observed ($I_{ph}$), making it a promising candidate for future technological applications.

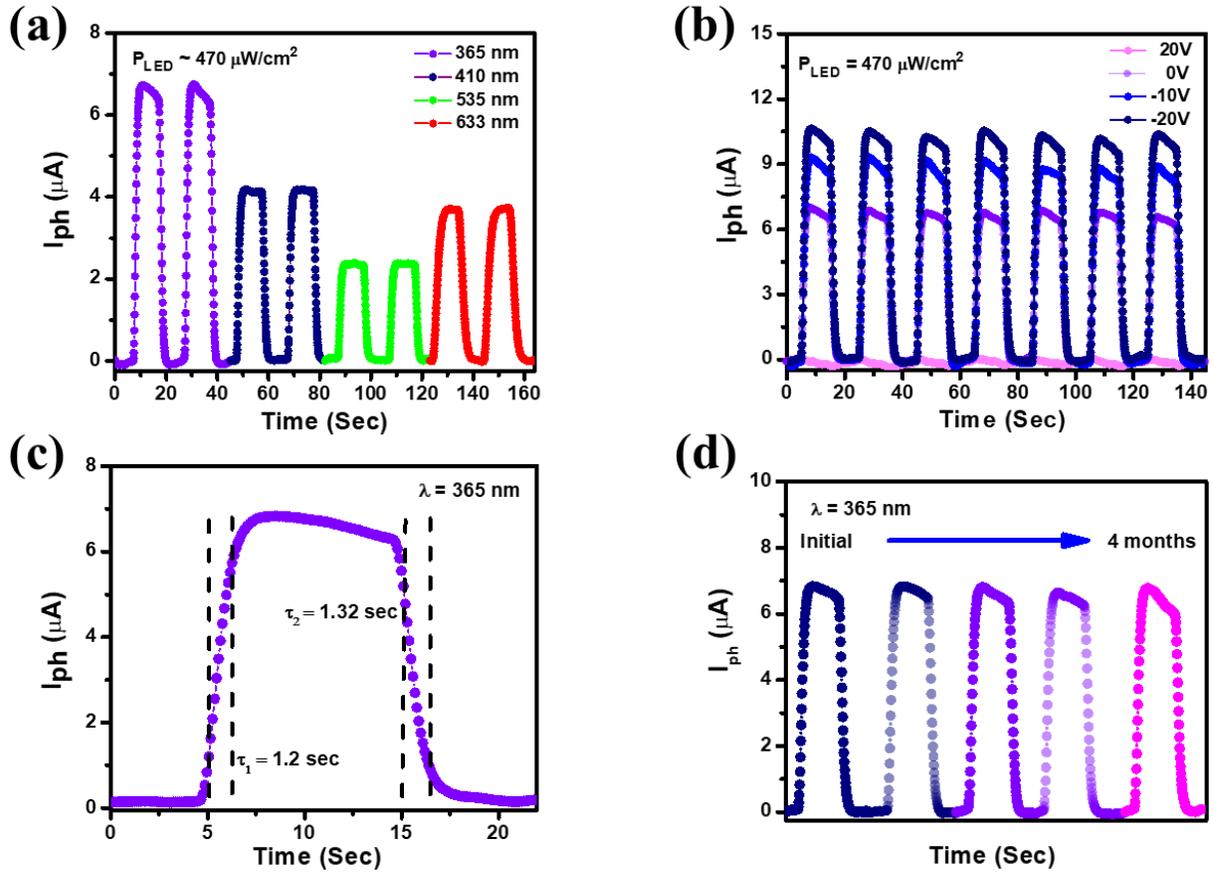

**Figure 4:** **(a)** Temporal photoresponse of the device with different wavelengths (365-633 nm) at $V_{ds}$ = 1V, $V_{bg}$ = 0V and $P_{LED}$ ~ 470 μW/cm² **(b)** Back gate ($V_{bg}$) dependent temporal photoresponse of the device under 365nm illumination **(c)** Enlarged part of **Figure 4 (a)** for 365 nm which shows the rise time ($\tau_1$) as 1.2 sec and the decay time ($\tau_2$) as 1.32 sec. **(d)** The temporal photoresponse after several months (4 months) of the device fabrication.

To get a deeper insight about the role of WS$_2$ QDs for high sensitive photodetection, a comparative study of the photoresponse is carried out on graphene-WS$_2$ phototransistors with two different morphologies of WS$_2$, viz., WS$_2$ quantum dots (0D) and WS$_2$ nanosheets (2D) (for 2D WS$_2$ nanosheets characterizations see **Figure S1**). All these devices are fabricated and measured under the same conditions. HRTEM and AFM images of WS$_2$ QDs and WS$_2$ nanosheets are shown in **Figure 5a, 5b** and **Figure S1c**, respectively. The temporal photocurrent and the responsivity for these two different devices are compared in **Figure 5c** and **5d,** respectively. We observe a significantly higher photoresponse (3.1×10² A/W) in Graphene -WS$_2$ QDs heterostructure than the Graphene -WS$_2$ nanosheets (1.83 × 10² A/W), suggesting that the morphology of WS$_2$ plays a major role in determining photocurrent. The photoresponse of a device strongly depends on the absorbance of the photo absorbing material and here it is experimentally observed that the absorbance is higher for WS$_2$ QDs than WS$_2$ nanosheets (**Figure S1b**).

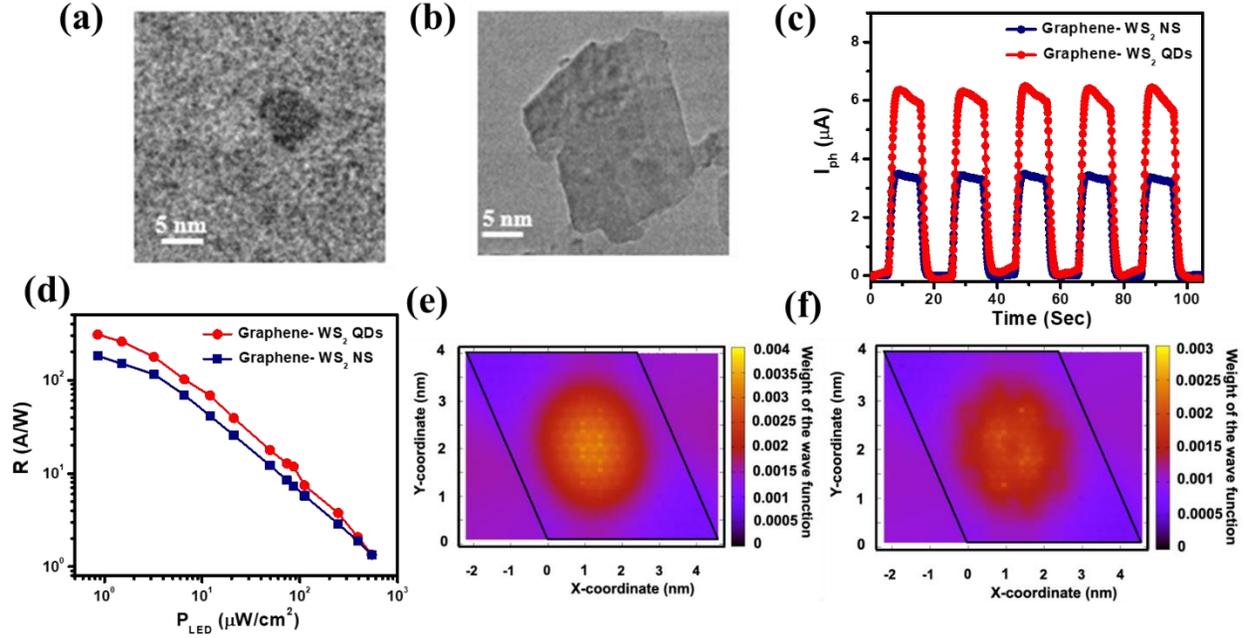

**Figure 5:** **(a)** and **(b)** represent the HRTEM images of WS$_2$ quantum dots and nanosheets respectively **(c)** Comparison between corresponding temporal photocurrents and **(d)** responsivities at 365 nm with $V_{ds}$ = 1V and $V_{bg}$ = 0V **(e)** Colour plot of the weight of the wave function corresponding to the highest occupied valence band on W atoms of WS$_2$ QDs with lateral dimension 4.73 nm **(f)** Colour plot of the weight of the wave function corresponding to the lowest unoccupied conduction band on W atoms of WS$_2$ QDs with a lateral dimension ~ 4.73 nm.

In order to understand the origin of enhanced photocurrent for WS$_2$ quantum dots (QDs), ab-initio electronic structure calculations are performed simulating a quantum dot (QD) with a size similar to our experimental result (~ 4.7 nm) (DOS, **Figure S5**). The calculated weight of the wave functions corresponding to the highest occupied valence band as well as the lowest unoccupied conduction band on each W atoms in a layer of the QD have been color-coded and are shown in **Figure 5e** and **5f**, respectively. For both these states, it has been found that the weight is not uniformly distributed over all the W atoms. These wave functions have higher weight at the center of the quantum dot, in contradiction to the case of infinite WS$_2$ nanosheet where the weight is uniformly distributed over all the W atoms and has a value 0. 0015 at a W site for the valence band maximum and 0.0011 for the conduction band minimum. One finds that oscillator strength for the transition will involve the weight of the valence band maximum wave function on a layer as well as the conduction band minimum wave function. This as we can see from **Figure 5e** leads to an enhancement of 2-3 times over what we have for the nanosheet. Similarly, **Figure 5f** reveals an enhancement of 2-3 times for the QD over the nanosheet. Consequently, one has an enhancement of around 9 times the oscillator strength for the QD over the nanosheets.

## **Conclusions**

In summary, we have fabricated a scalable, eco-friendly, cost effective, mixed dimensional (2D-0D) hybrid phototransistor by using a CVD grown graphene and chemically exfoliated WS$_2$ quantum dots without any lithographic process. Semiconducting WS$_2$ QDs (average radius ~ 2.75 nm) are synthesized by simple chemical exfoliation procedure, which act as a photo-absorbing material in the heterostructure device. Raman and photoluminescence measurements indicate the charge transfer between WS$_2$ QDs and the graphene layer, and the fabricated device shows an excellent UV-visible broadband (365-633 nm) photo response with superior gate tunability. The hybrid device exhibits high detectivity (~$2.2\times10^{12}$ Jones), superior photoresponsivity ($3.3\times10^2$ A/W) and low noise equivalent power (~$4\times10^{-14}$ W/Hz$^{0.5}$), which are comparable or better than other available graphene based hybrid photodetectors[55]. The response time of ~1.2 sec (rise time) and ~1.32 sec (fall time) are also comparable or faster than the previously reported graphene based hybrid photodetectors. The solution processed 0D/2D phototransistor device exhibits an extraordinary stability without any degradation of photocurrent even after several months of its fabrication, demonstrating its potential for future optoelectronic devices. Furthermore, a comparative study has been done between two different morphologies of WS$_2$ such as graphene-WS$_2$ quantum dots (2D-0D) heterostructure and graphene-WS$_2$ nanosheets (2D-2D) heterostructure with the former device showing better photoresponse, which can be attributed to the higher absorbance of WS$_2$ QDs due to the quantum confinement effect. This is further explained by our first principle calculation using DFT. This superior quality large area, broadband, hybrid phototransistor has significant potential in the applications of imaging, sensing, and communication etc. The obtained results in this work demonstrate that the solution processed wafer scale WS$_2$ QDs based hybrid device to be a prospective candidate for the next generation low cost, high performance optoelectronic devices.

**Table 1. Comparison of performances with reported graphene-TMDC hybrid photodetectors**

| SL No | Device structure | Channel length (μm) | Wavelength (nm) | $V_{sd}$ (V) | $V_{bg}$ (V) | R (A/W) | D* (Jones) | $\tau_{Rise}/\tau_{Fall}$ (Sec) | references |
|---|---|---|---|---|---|---|---|---|---|
| 1. | Graphene-MoS$_2$ | ~ 20 | 632.8 | 0.1 | -20 | 10 | - | 0.280/1.50 | 55 |
| 2. | Graphene-WS$_2$* | 250 | 550 | 5 | 0 | 6.4 | $2.8\times10^{10}$ | 0.01/0.02 | 41 |
| 3. | Graphene-WS$_2$ | - | 532 | 5 | 30 | 121 | - | - | 56 |
| 4. | Graphene-MoTe$_2$ | ~10 | 532 | -2 | 0 | 0.02 | - | 0.03/0.03 | 57 |
| 5. | Graphene-MoS$_2$-graphene | ~5 | 405-2000 | -1 | 40 | 414 | $3.2\times10^{10}$ | 0.692/ - | 63 |
| 6. | Graphene-WS$_2$ QDs * | 1000 | 365 | 5 | 0 | 1814 | $7.47\times10^{12}$ | 2.04/2.89 | 66 |
| 7. | Graphene-WS$_2$-Al$_2$O$_3$ | 90 | 340-680 | 2 | 0 | 950 | - | 7.85/5.61 | 64 |
| 8. | Graphene-BP | ~ 3 | 655-980 | 0.5 | -35 | 55.75 | - | -/0.036 | 59 |
| 9. | Graphene-WS$_2$ QDs | ~500 | 365-633 | 1 | 0 | $3.1\times10^2$ | $2.2\times10^{12}$ | 1.2/1.32 | **This work** |

*2 terminal diode configuration

## Materials and methods

### Synthesis of WS$_2$ QDs

$WS_2$ QDs are synthesized chemically by using Li-intercalation technique. First, for exfoliation of $WS_2$ nanosheets, bulk $WS_2$ powder 2.5 gm with anhydrous LiBr at 1:1 molar ratio are dispersed in 25 ml hexane solution. Then this solution is sonicated for 5 hrs by using a bath sonicator. After sonication, the resulting black dispersion is centrifuged at 5000 rpm for 10 mins to remove hexane and untreated Li ions. The wet sediments are washed by dispersing in IPA by shaking followed by centrifugation (5000 rpm, 15 mins). By repeating this procedure three times the wet sediment of $WS_2$ is completely transferred in IPA solvent. Then after 1hr bath sonication, the resulting dispersion is centrifuged at 5000 rpm for 10 mins resulting a greenish colored exfoliated $WS_2$ nanosheets. To get the $WS_2$ QDs, the dispersion of $WS_2$ nanosheets in IPA is bath sonicated further for 10 hrs to achieve fragmented quantum dots (QDs), due to prolonged sonication, the lateral dimension of the $WS_2$ nanosheets are reduced in all directions, resulting in formation of 0D QDs. Then the resulting dispersion is centrifuged at 10000 rpm for 30 mins and the supernatant is collected which is the dispersion of $WS_2$ QDs in IPA.

**Materials characterizations**

The phase and crystallinity of synthesized $WS_2$ quantum dots (QDs) are investigated by X-ray diffraction (PANalytical X-PERT PRO) using Cu-K$\alpha$ radiation (1.54 Å). Surface profile of the hybrid heterostructure is examined by using a field-emission scanning electron microscope (FE-SEM, QUANTA FEG 250) with an electron energy of 20 keV and equipped with an energy-dispersive X-ray (EDX) spectrometer. The morphology of chemically exfoliated $WS_2$ QDs and NS are examined by using a high-resolution transmission electron microscope (FEI-TECNAI G2 20ST, energy 200 keV) and atomic force microscopy (di INNOVA). Absorption spectrum of chemically exfoliated $WS_2$ QDs and NS are measured using UV-Vis spectrometer (Shimadzu - UV-Vis 2600 Spectrophotometer) and Raman and Photoluminescence spectrum are recorded in LabRam HR Evolution; HORIBA France SAS-532 nm laser.

**Fabrication of Graphene-$WS_2$ heterostructures**

The phototransistor devices are fabricated by using commercially available single-layer CVD-grown graphene on p+ doped $Si/SiO_2$ (300 nm) substrates (purchased from Graphenea, USA). The electrical contacts are made by simply putting two silver epoxy blobs on the graphene sheet. Before measurements, the edges of the graphene sheet are mechanically removed to make sure there is no electrical contact between the graphene layer and the Si backgate. The channel length (L) and the width (W) are taken as ~500 μm and ~1500 μm respectively. After that, $WS_2$ QDs solution (8 μl) is drop casted onto the graphene FET to fabricate the heterostructure. Finally, after keeping in ambient condition for 30 mins the hybrid device is dried at 80°C for 1 hr. in order to get better interfacing.

**Optoelectronic measurements**

All the electrical and optical measurements are carried out at room temperature (in vacuum $\sim 10^{-5}$ mbar) in a homemade electronic setup having an optical window. MFLI lock in amplifier (Zurich Instrument) and a Keithley 2450 sourcemeter are used in all over the experiments with AC two-probe configuration at a carrier frequency of 83.6 Hz. For the photocurrent measurements we use well calibrated and collimated 365 nm, 410 nm, 535 nm and 633 nm LEDs (Thorlab) sourced by DC2200 power supply with a spot size ~ 3 mm. The LED powers are calibrated by using Flame-Ocean Optics spectrometer with integrating sphere set up.

**Methods for DFT Simulation**

Ab-initio electronic structure calculations have been carried out for a periodic bilayer of $WS_2$ using density functional theory as implemented in Vienna ab initio simulation package (VASP). Projector augmented plane wave (PAW) potentials have been used with a plane-wave basis set implementation. We have used the generalized gradient approximation as given by Perdew- Burke-Ernzerhof (PBE) functional form for the exchange correlation functional. The lattice parameters were kept fixed at of 0.315 nm while the internal atomic positions were relaxed by minimizing the forces on each atom. A vacuum of 2 nm has been incorporated along the z-direction to minimize interactions between the periodic images inevitably present in the implementation of periodic unit cells that we use. Van der Waals interactions were included between the layers of $WS_2$, using DFT-D2 method of Grimme. This led to an interlayer separation of 0.29 nm. A Gamma centered k-point mesh of 12x12x1 has been used for the k-point integrations in the calculations. The ab initio band structure calculated for bilayer $WS_2$ was then mapped onto a tight binding model to extract the onsite energies and the hopping interaction strengths, using a least square error minimization method. The tight-binding model considered included s, p and d states on both W and S atoms and have been parameterized in terms of the Slater Koster interaction strengths. This was then used to calculate the electronic structure of the bilayer nanoflake of dimension 4.73 nm for which the density of states are given in **Figure S5**. The surface atoms in the nanoflake were passivated with pseudo atoms, and the contribution of the surface atoms was removed in order to identify the valence band maximum and conduction band minimum.

## Acknowledgement

S.M. acknowledges the support from DST-Inspire fellowship. S.P. acknowledges SERB, Govt. of India for fellowship. P. M. acknowledges support from DST - Nanomission for the research through the project DST/NM/NS/2018/18. We also acknowledge the characterization facilities from the TRC project, SNBNCBS.

## References

(1) Yao, J.; Yang, G. 2D Material Broadband Photodetectors. *Nanoscale* **2020**, *12* (2), 454–476. https://doi.org/10.1039/c9nr09070c.
(2) Jariwala, D.; Marks, T. J.; Hersam, M. C. Mixed-Dimensional van Der Waals Heterostructures. *Nat. Mater.* **2017**, *16* (2), 170–181. https://doi.org/10.1038/nmat4703.
(3) Koppens, F. H. L.; Mueller, T.; Avouris, P.; Ferrari, A. C.; Vitiello, M. S.; Polini, M. Photodetectors Based on Graphene, Other Two-Dimensional Materials and Hybrid Systems. *Nat. Nanotechnol.* **2014**, *9* (10), 780–793. https://doi.org/10.1038/nnano.2014.215.
(4) Liu, K.; Sakurai, M.; Aono, M. ZnO-Based Ultraviolet Photodetectors. *Sensors* **2010**, *10* (9), 8604–8634. https://doi.org/10.3390/s100908604.
(5) Velazquez, R.; Aldalbahi, A.; Rivera, M.; Feng, P. Fabrications and Application of Single Crystalline GaN for High-Performance Deep UV Photodetectors. *AIP Adv.* **2016**, *6* (8).

https://doi.org/10.1063/1.4961878.

(6) Das, K.; Mukherjee, S.; Manna, S.; Ray, S. K.; Raychaudhuri, A. K. Single Si Nanowire (Diameter ≤ 100 Nm) Based Polarization Sensitive near-Infrared Photodetector with Ultra-High Responsivity. *Nanoscale* **2014**, *6* (19), 11232–11239. https://doi.org/10.1039/c4nr03170a.

(7) Ray, S. K.; Katiyar, A. K.; Raychaudhuri, A. K. One-Dimensional Si/Ge Nanowires and Their Heterostructures for Multifunctional Applications - A Review. *Nanotechnology* **2017**, *28* (9). https://doi.org/10.1088/1361-6528/aa565c.

(8) Clifford, J. P.; Konstantatos, G.; Johnston, K. W.; Hoogland, S.; Levina, L.; Sargent, E. H. Fast, Sensitive and Spectrally Tuneable Colloidal-Quantum-Dot Photodetectors. *Nat. Nanotechnol.* **2009**, *4* (1), 40–44. https://doi.org/10.1038/nnano.2008.313.

(9) Tan, C. L.; Mohseni, H. Emerging Technologies for High Performance Infrared Detectors. *Nanophotonics* **2018**, *7* (1), 169–197. https://doi.org/10.1515/nanoph-2017-0061.

(10) Novoselov, K. S.; Geim, A. K.; Morozov, S. V.; Jiang, D.; Katsnelson, M. I.; Grigorieva, I. V.; Dubonos, S. V.; Firsov, A. A. Two-Dimensional Gas of Massless Dirac Fermions in Graphene. *Nature* **2005**, *438* (7065), 197–200. https://doi.org/10.1038/nature04233.

(11) Radisavljevic, B.; Radenovic, A.; Brivio, J.; Giacometti, V.; Kis, A. Single-Layer MoS2 Transistors. *Nat. Nanotechnol.* **2011**, *6* (3), 147–150. https://doi.org/10.1038/nnano.2010.279.

(12) Morozov, S. V.; Novoselov, K. S.; Katsnelson, M. I.; Schedin, F.; Elias, D. C.; Jaszczak, J. A.; Geim, A. K. Giant Intrinsic Carrier Mobilities in Graphene and Its Bilayer. *Phys. Rev. Lett.* **2008**, *100* (1), 11–14. https://doi.org/10.1103/PhysRevLett.100.016602.

(13) Novoselov, K. S.; Fal'Ko, V. I.; Colombo, L.; Gellert, P. R.; Schwab, M. G.; Kim, K. A Roadmap for Graphene. *Nature* **2012**, *490* (7419), 192–200. https://doi.org/10.1038/nature11458.

(14) Bonaccorso, F.; Sun, Z.; Hasan, T.; Ferrari, A. C. Graphene Photonics and Optoelectronics. *Nat. Photonics* **2010**, *4* (9), 611–622. https://doi.org/10.1038/nphoton.2010.186.

(15) Fang, Z.; Wang, Y.; Liu, Z.; Schlather, A.; Ajayan, P. M.; Koppens, F. H. L.; Nordlander, P.; Halas, N. J. Plasmon-Induced Doping of Graphene. *ACS Nano* **2012**, *6* (11), 10222–10228. https://doi.org/10.1021/nn304028b.

(16) Furchi, M.; Urich, A.; Pospischil, A.; Lilley, G.; Unterrainer, K.; Detz, H.; Klang, P.; Andrews, A. M.; Schrenk, W.; Strasser, G.; Mueller, T. Microcavity-Integrated Graphene Photodetector. *Nano Lett.* **2012**, *12* (6), 2773–2777. https://doi.org/10.1021/nl204512x.

(17) Konstantatos, G.; Badioli, M.; Gaudreau, L.; Osmond, J.; Bernechea, M.; De Arquer, F. P. G.; Gatti, F.; Koppens, F. H. L. Hybrid Grapheneĝquantum Dot Phototransistors with Ultrahigh Gain. *Nat. Nanotechnol.* **2012**, *7* (6), 363–368. https://doi.org/10.1038/nnano.2012.60.

(18) Gong, M.; Liu, Q.; Cook, B.; Kattel, B.; Wang, T.; Chan, W. L.; Ewing, D.; Casper, M.; Stramel, A.; Wu, J. Z. All-Printable ZnO Quantum Dots/Graphene van Der Waals Heterostructures for Ultrasensitive Detection of Ultraviolet Light. *ACS Nano* **2017**, *11* (4), 4114–4123. https://doi.org/10.1021/acsnano.7b00805.

(19) Ni, Z.; Ma, L.; Du, S.; Xu, Y.; Yuan, M.; Fang, H.; Wang, Z.; Xu, M.; Li, D.; Yang, J.; Hu, W.; Pi, X.; Yang, D. Plasmonic Silicon Quantum Dots Enabled High-Sensitivity Ultrabroadband Photodetection of Graphene-Based Hybrid Phototransistors. *ACS Nano* **2017**, *11* (10), 9854–9862. https://doi.org/10.1021/acsnano.7b03569.

(20) Roy, K.; Padmanabhan, M.; Goswami, S.; Sai, T. P.; Ramalingam, G.; Raghavan, S.; Ghosh, A. Graphene-MoS 2 Hybrid Structures for Multifunctional Photoresponsive Memory Devices. *Nat. Nanotechnol.* **2013**, *8* (11), 826–830. https://doi.org/10.1038/nnano.2013.206.

(21) Mehew, J. D.; Unal, S.; Torres Alonso, E.; Jones, G. F.; Fadhil Ramadhan, S.; Craciun, M. F.; Russo, S. Fast and Highly Sensitive Ionic-Polymer-Gated WS2–Graphene Photodetectors. *Adv. Mater.* **2017**, *29* (23). https://doi.org/10.1002/adma.201700222.

(22) Sun, Z.; Liu, Z.; Li, J.; Tai, G.; Lau, S.; Yan, F. Infrared Photodetectors Based on CVD-Grown Graphene and PbS Quantum Dots with Ultrahigh Responsivity. **2012**, 1–6. https://doi.org/10.1002/adma.201202220.

(23) Spirito, D.; Kudera, S.; Miseikis, V.; Giansante, C.; Coletti, C.; Krahne, R. UV Light Detection from

(23) CdS Nanocrystal Sensitized Graphene Photodetectors at KHz Frequencies. **2015**. https://doi.org/10.1021/acs.jpcc.5b07895.

(24) Shao, Y.; Liu, Y.; Chen, X.; Chen, C.; Sarpkaya, I.; Chen, Z.; Fang, Y.; Kong, J.; Watanabe, K.; Taniguchi, T.; Taylor, A.; Huang, J.; Xia, F. Stable Graphene-Two-Dimensional Multiphase Perovskite Heterostructure Phototransistors with High Gain. *Nano Lett.* **2017**, *17* (12), 7330–7338. https://doi.org/10.1021/acs.nanolett.7b02980.

(25) Pradhan, B.; Das, S.; Li, J.; Chowdhury, F.; Cherusseri, J.; Pandey, D.; Dev, D.; Krishnaprasad, A.; Barrios, E.; Towers, A.; Gesquiere, A.; Tetard, L.; Roy, T.; Thomas, J. Ultrasensitive and Ultrathin Phototransistors and Photonic Synapses Using Perovskite Quantum Dots Grown from Graphene Lattice. *Sci. Adv.* **2020**, *6* (7), 1–12. https://doi.org/10.1126/sciadv.aay5225.

(26) Eda, G.; Yamaguchi, H.; Voiry, D.; Fujita, T.; Chen, M.; Chhowalla, M. Photoluminescence from Chemically Exfoliated MoS 2. **2011**, 5111–5116. https://doi.org/10.1021/nl201874w.

(27) Nicolosi, V.; Chhowalla, M.; Kanatzidis, M. G.; Strano, M. S.; Coleman, J. N. Liquid Exfoliation of Layered Materials. *Science (80-. ).* **2013**, *340* (6139), 72–75. https://doi.org/10.1126/science.1226419.

(28) Ghorai, A.; Bayan, S.; Gogurla, N.; Midya, A.; Ray, S. K. Highly Luminescent WS 2 Quantum Dots/ZnO Heterojunctions for Light Emitting Devices. **2017**. https://doi.org/10.1021/acsami.6b12859.

(29) Ghorai, A.; Midya, A.; Maiti, R.; Ray, S. K. Exfoliation of WS2 in the Semiconducting Phase Using a Group of Lithium Halides: A New Method of Li Intercalation. *Dalt. Trans.* **2016**, *45* (38), 14979–14987. https://doi.org/10.1039/c6dt02823c.

(30) Nanocomposites, L. W. S.; Rout, C. S.; Joshi, P. D.; Kashid, R. V; Joag, D. S.; More, M. A.; Simbeck, A. J.; Washington, M.; Nayak, S. K.; Late, D. J. Superior Field Emission Properties Of. **2013**, 1–8. https://doi.org/10.1038/srep03282.

(31) Yin, W.; Bai, X.; Chen, P.; Zhang, X.; Su, L.; Ji, C.; Gao, H. Rational Control of Size and Photoluminescence of WS 2 Quantum Dots for White Light-Emitting Diodes. **2018**. https://doi.org/10.1021/acsami.8b17966.

(32) Thin, A. Evolution of Electronic Structure In. **2013**, No. 1, 791–797. https://doi.org/10.1021/nn305275h.

(33) Anupama, S.; Kaul, B. As Featured in : Enhancement in Environmentally-Friendly Solution. *J. Mater. Chem. C* **2017**, *5*, 5323–5333. https://doi.org/10.1039/c7tc01001j.

(34) Carvalho, A.; Ribeiro, R. M.; Neto, A. H. C. Band Nesting and the Optical Response of Two-Dimensional Semiconducting Transition Metal Dichalcogenides. **2013**, *115205*, 1–6. https://doi.org/10.1103/PhysRevB.88.115205.

(35) Wang, L.; Wang, Z.; Wang, H.; Grinblat, G.; Huang, Y.; Wang, D.; Ye, X.; Li, X.; Bao, Q.; Wee, A.; Maier, S. A.; Chen, Q.; Zhong, M. Hot Carriers in MoS 2 Monolayer. **2017**. https://doi.org/10.1038/ncomms13906.

(36) Bayat, A.; Saievar-iranizad, E. Synthesis of Blue Photoluminescent WS 2 Quantum Dots via Ultrasonic Cavitation. *J. Lumin.* **2017**, *185*, 236–240. https://doi.org/10.1016/j.jlumin.2017.01.024.

(37) Liu, P.; Liu, Y.; Ye, W.; Ma, J.; Gao, D. Flower-like N-Doped MoS 2 for Photocatalytic Degradation of RhB by Visible Light Irradiation. https://doi.org/10.1088/0957-4484/27/22/225403.

(38) McCreary, A.; Berkdemir, A.; Wang, J.; Nguyen, M. A.; Elías, A. L.; Perea-López, N.; Fujisawa, K.; Kabius, B.; Carozo, V.; Cullen, D. A.; Mallouk, T. E.; Zhu, J.; Terrones, M. Distinct Photoluminescence and Raman Spectroscopy Signatures for Identifying Highly Crystalline WS2 Monolayers Produced by Different Growth Methods. *J. Mater. Res.* **2016**, *31* (7), 931–944. https://doi.org/10.1557/jmr.2016.47.

(39) Lan, C.; Li, C.; Wang, S.; He, T.; Jiao, T.; Wei, D.; Jing, W.; Li, L.; Liu, Y. Zener Tunneling and Photoresponse of a WS2/Si van Der Waals Heterojunction. *ACS Appl. Mater. Interfaces* **2016**, *8* (28), 18375–18382. https://doi.org/10.1021/acsami.6b05109.

(40) Ferrari, A. C.; Basko, D. M. Studying the Properties of Graphene. *Nat. Publ. Gr.* **2013**, *8* (April), 235–246. https://doi.org/10.1038/nnano.2013.46.

(41) Alamri, M.; Gong, M.; Cook, B.; Goul, R.; Wu, J. Z. Plasmonic WS2 Nanodiscs/Graphene van Der Waals Heterostructure Photodetectors. *ACS Appl. Mater. Interfaces* **2019**, *11* (36), 33390–33398. https://doi.org/10.1021/acsami.9b09262.

(42) Gong, M.; Sakidja, R.; Liu, Q.; Goul, R.; Ewing, D.; Casper, M.; Stramel, A.; Elliot, A.; Wu, J. Z. Broadband Photodetectors Enabled by Localized Surface Plasmonic Resonance in Doped Iron Pyrite Nanocrystals. *Adv. Opt. Mater.* **2018**, *6* (8), 1–11. https://doi.org/10.1002/adom.201701241.

(43) Luther, J. M.; Jain, P. K.; Ewers, T.; Alivisatos, A. P. Localized Surface Plasmon Resonances Arising from Free Carriers in Doped Quantum Dots. *Nat. Mater.* **2011**, *10* (5), 361–366. https://doi.org/10.1038/nmat3004.

(44) Notley, S. M. High Yield Production of Photoluminescent Tungsten Disulphide Nanoparticles. *Journal of Colloid and Interface Science*. 2013, pp 160–164. https://doi.org/10.1016/j.jcis.2013.01.035.

(45) Lin, W.; Zhuang, P.; Chou, H.; Gu, Y.; Roberts, R.; Li, W.; Banerjee, S. K.; Cai, W.; Akinwande, D. Electron Redistribution and Energy Transfer in Graphene/MoS2 Heterostructure. *Appl. Phys. Lett.* **2019**, *114* (11). https://doi.org/10.1063/1.5088512.

(46) Wang, Y.; Zhang, Y.; Lu, Y.; Xu, W.; Mu, H.; Chen, C.; Qiao, H.; Song, J.; Li, S.; Sun, B.; Cheng, Y. B.; Bao, Q. Hybrid Graphene-Perovskite Phototransistors with Ultrahigh Responsivity and Gain. *Adv. Opt. Mater.* **2015**, *3* (10), 1389–1396. https://doi.org/10.1002/adom.201500150.

(47) Kim, S.; Shin, S.; Kim, T.; Du, H.; Song, M.; Lee, C. W.; Kim, K.; Cho, S.; Seo, D. H.; Seo, S. Robust Graphene Wet Transfer Process through Low Molecular Weight Polymethylmethacrylate. *Carbon N. Y.* **2016**, *98*, 352–357. https://doi.org/10.1016/j.carbon.2015.11.027.

(48) Pal, A. N.; Ghosh, A. Ultralow Noise Field-Effect Transistor from Multilayer Graphene. *Appl. Phys. Lett.* **2009**, *95* (8), 8–10. https://doi.org/10.1063/1.3206658.

(49) Soci, C.; Zhang, A.; Xiang, B.; Dayeh, S. A.; Aplin, D. P. R.; Park, J.; Bao, X. Y.; Lo, Y. H.; Wang, D. ZnO Nanowire UV Photodetectors with High Internal Gain. **2007**, 1–7. https://doi.org/10.1021/nl070111x.

(50) Liu, Y.; Wang, F.; Wang, X.; Wang, X.; Flahaut, E.; Liu, X.; Li, Y.; Wang, X.; Xu, Y.; Shi, Y.; Zhang, R. Planar Carbon Nanotube–Graphene Hybrid Films for High-Performance Broadband Photodetectors. *Nat. Commun.* **2015**, 1–7. https://doi.org/10.1038/ncomms9589.

(51) Liu, X.; Luo, X.; Nan, H.; Guo, H.; Wang, P.; Zhang, L.; Xu, J.; Wang, X. Epitaxial Ultrathin Organic Crystals on Graphene for High- Effi Ciency Phototransistors. **2016**, 5200–5205. https://doi.org/10.1002/adma.201600400.

(52) Ahmad, S.; Kanaujia, P. K.; Beeson, H. J.; Abate, A.; Deschler, F.; Credgington, D.; Steiner, U.; Prakash, G. V.; Baumberg, J. J. Strong Photocurrent from Two-Dimensional Excitons in Solution-Processed Stacked Perovskite Semiconductor Sheets. *ACS Appl. Mater. Interfaces* **2015**, *7* (45), 25227–25236. https://doi.org/10.1021/acsami.5b07026.

(53) Lan, C.; Li, C.; Yin, Y.; Liu, Y. Correction: Large-Area Synthesis of Monolayer WS2 and Its Ambient-Sensitive Photo-Detecting Performance. **2015**, *4* (d), 610054. https://doi.org/10.1039/c5nr90175h.

(54) Mueller, T.; Xia, F.; Avouris, P. Graphene Photodetectors for High-Speed Optical Communications. *Nat. Photonics* **2010**, *4* (5), 297–301. https://doi.org/10.1038/nphoton.2010.40.

(55) Xu, H.; Wu, J.; Feng, Q.; Mao, N.; Wang, C. High Responsivity and Gate Tunable Graphene-MoS 2 Hybrid Phototransistor. **2014**, 1–7. https://doi.org/10.1002/smll.201303670.

(56) Chen, T.; Sheng, Y.; Zhou, Y.; Chang, R. J.; Wang, X.; Huang, H.; Zhang, Q.; Hou, L.; Warner, J. H. High Photoresponsivity in Ultrathin 2D Lateral Graphene:WS 2 :Graphene Photodetectors Using Direct CVD Growth. *ACS Appl. Mater. Interfaces* **2019**. https://doi.org/10.1021/acsami.8b20321.

(57) Phys, A. Enhancing Photoresponsivity Using MoTe 2 - Graphene Vertical Heterostructures. **2016**, *063506* (October 2015). https://doi.org/10.1063/1.4941996.

(58) Xiao, R.; Lan, C.; Li, Y.; Zeng, C.; He, T.; Wang, S. High Performance Van Der Waals Graphene – WS 2 – Si Heterostructure Photodetector. **2019**, *1901304*, 1–7. https://doi.org/10.1002/admi.201901304.

(59) Xu, J.; Song, Y. J.; Park, J. H.; Lee, S. Graphene/Black Phosphorus Heterostructured Photodetector. *Solid. State. Electron.* **2018**, *144* (January), 86–89. https://doi.org/10.1016/j.sse.2018.03.007.
(60) Xie, C.; Mak, C.; Tao, X.; Yan, F. Photodetectors Based on Two-Dimensional Layered Materials Beyond Graphene. *Adv. Funct. Mater.* **2017**, *27* (19). https://doi.org/10.1002/adfm.201603886.
(61) Mukherjee, S.; Das, K.; Das, S.; Ray, S. K. Highly Responsive, Polarization Sensitive, Self-Biased Single $GeO_2$ - Ge Nanowire Device for Broadband and Low Power Photodetectors. **2018**. https://doi.org/10.1021/acsphotonics.8b00981.
(62) Islam, S.; Mishra, J. K.; Kumar, A.; Chatterjee, D.; Ravishankar, N.; Ghosh, A. Ultra-Sensitive Graphene-Bismuth Telluride Nano-Wire Hybrids for Infrared Detection. *Nanoscale* **2019**, *11* (4), 1579–1586. https://doi.org/10.1039/c8nr08433e.
(63) Gao, S.; Wang, Z.; Wang, H.; Meng, F.; Wang, P.; Chen, S.; Zeng, Y.; Zhao, J.; Hu, H.; Cao, R.; Xu, Z.; Guo, Z. Graphene / $MoS_2$ / Graphene Vertical Heterostructure-Based Broadband Photodetector with High Performance. **2020**, *2001730*, 1–6. https://doi.org/10.1002/admi.202001730.
(64) Changyong Lan, Y. L. High Responsive and Broadband Photodetectors Based on WS2-Graphene van Der Waals Epitaxial Heterostructures. *J. Mater. Chem. C* **2017**. https://doi.org/10.1039/C6TC05037A.
(65) Ago, H.; Endo, H.; Sol, P. Controlled van Der Waals Epitaxy of Monolayer $MoS_2$ Triangular Domains on Graphene. **2015**. https://doi.org/10.1021/am508569m.
(66) Pandey, A.; Srivastava, A. WS2 Quantum Dot Graphene Nanocomposite Film for UV Photodetection. *ACS Appl. Nano Mater.* **2019**, *2*, 3934–3942. https://doi.org/10.1021/acsanm.9b00820.

# Supporting Information

**High Responsivity Gate Tunable UV-Visible Broadband Phototransistor Based on Graphene –WS$_2$ Mixed Dimensional (2D-0D) Heterostructure**

Shubhrasish Mukherjee[1], Didhiti Bhattacharya[1], Sumanti Patra[1], Sanjukta Paul[1], Rajib Kumar Mitra[1], Priya Mahadevan[1], Atindra Nath Pal*[1] and Samit Kumar Ray*[1,2]

[1]*S. N. Bose National Center for Basic Science, Sector III, Block JD, Salt Lake, Kolkata – 700106*

[2]*Indian Institute of Technology Kharagpur, 721302, West Bengal, India*

**\***Email: atin@bose.res.in**,** physkr@phy.iitkgp.ac.in

**Contents**

**Supplementary Note 1:** Characterizations of WS$_2$ quantum dots and nanosheets

**Supplementary Note 2:** Characterizations of the graphene-WS$_2$ QDs hybrid device

**Supplementary Note 3:** Calculation of carrier mobility of the graphene-WS$_2$ QDs hybrid device

**Supplementary Note 4:** Optoelectronics characterization of the graphene-WS$_2$ hybrid device

**Supplementary Note 5:** Responsivity, Noise equivalent power (NEP) and the specific detectivity (D*) of the graphene-WS$_2$ hybrid device

**Supplementary Note 6:** Density of states (DOS) calculation of WS$_2$ quantum dots (QDs)

1. **Characterizations of WS$_2$ quantum dots and nanosheets**

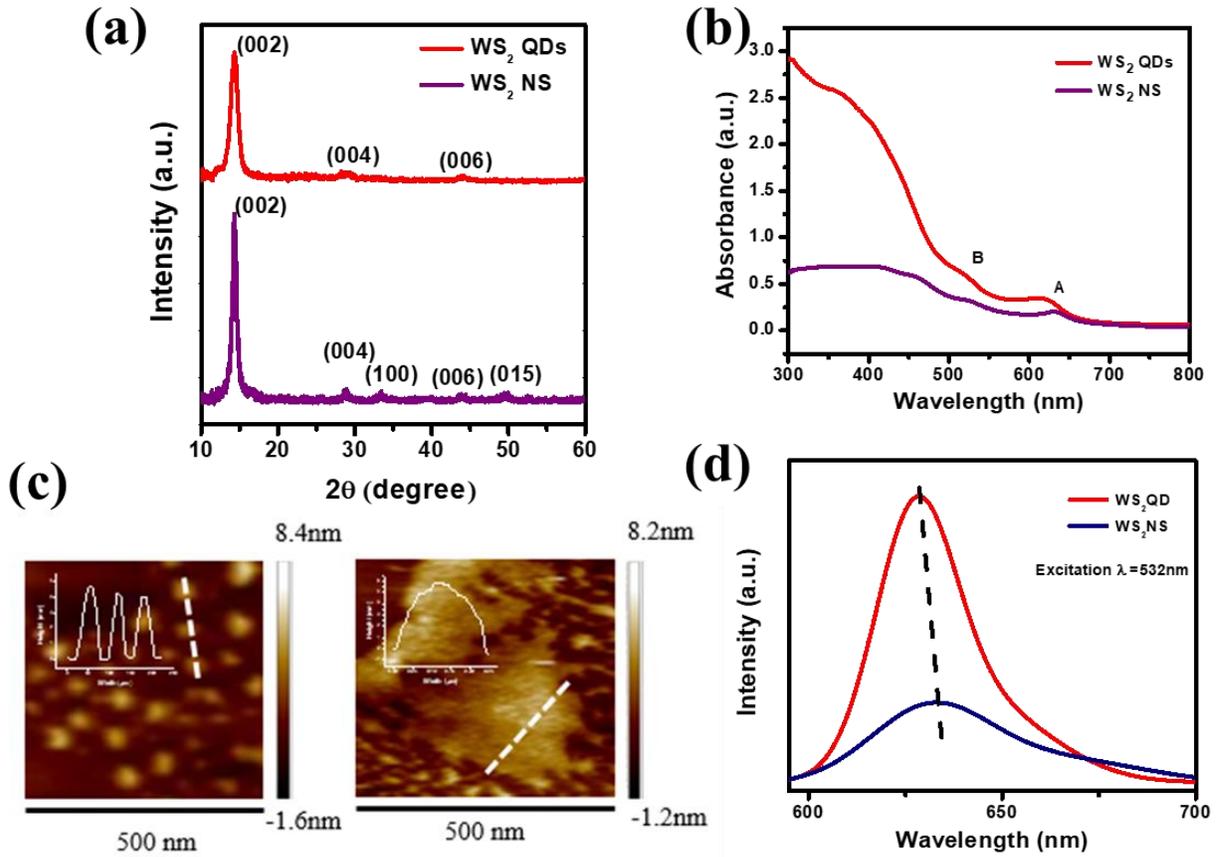

**Figure S 1(a)** XRD pattern of WS$_2$ quantum dots (QDs) and WS$_2$ nano-sheets (NS) **(b)** Absorbance spectra of WS$_2$ QDs and WS$_2$ NS **(c)** AFM images of WS$_2$ QDs (left) and WS$_2$ NS (right) **(d)** PL spectra of WS$_2$ QDs and WS$_2$ NS.

## 2. Characterizations of the graphene-WS$_2$ QDs hybrid device

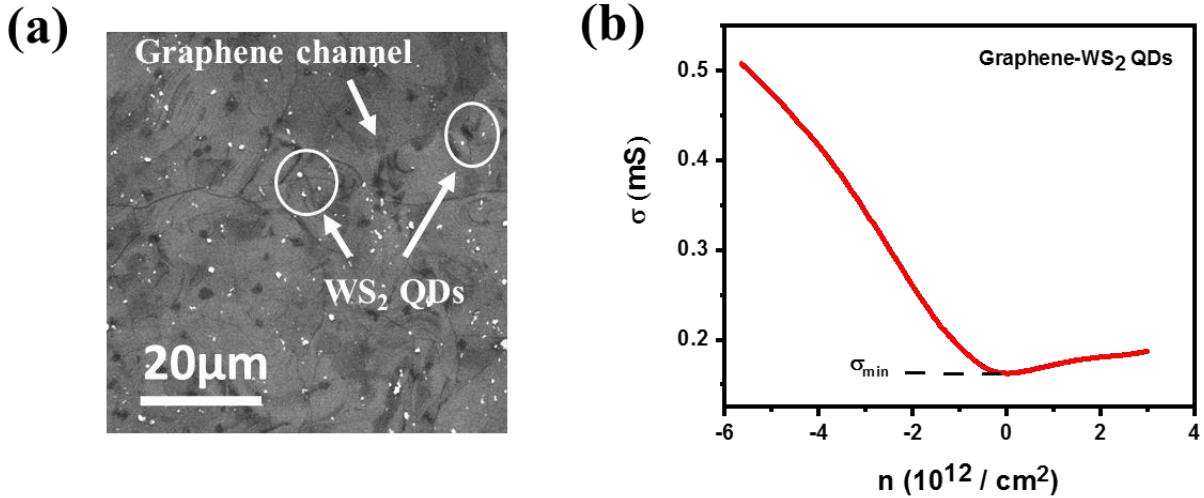

**Figure S 2(a)** SEM image of the hybrid graphene-WS$_2$ device **(b)** Conductivity (σ) as a function of density (n) of the hybrid device in dark.

## 3. Calculation of carrier mobility of the graphene-WS$_2$ QDs hybrid device

The field effect mobility (μ) of the hybrid graphene-WS$_2$ QDs transistor can be calculated by simply introducing Drude model which gives

$$\mu = \frac{\sigma}{ne} \quad \text{...............................} (1)$$

Where, σ is the conductivity, n represents the charge carrier density and e is the electronic charge.

Here, σ is calculated from the relation,

$$\sigma = \frac{1}{R}\frac{L}{W} \quad \text{..................................} (2)$$

Where, R is the resistance of the device. L and W are the effective length and width of the device Here in our device L~ 500 μm and W~ 1500 μm.

The gate voltage ($V_{bg}$) is converted to carrier density (n) using a parallel plate capacitor model, i.e.

$$n = \frac{C_{bg}}{e}(V_{bg} - V_D) \quad \text{.......................} (3)$$

Where, $C_{bg}$ = 1.15×10$^{-8}$ F/ cm$^2$ is the backgate capacitance of 300 nm SiO$_2$. Using the following equations we have calculated the field effect mobility of the charge carriers as ~ 400 cm$^2$/VS for holes and ~ 61 cm$^2$/VS for electrons in our hybrid device.

## 4. Optoelectronics characterization of the graphene-WS$_2$ hybrid device

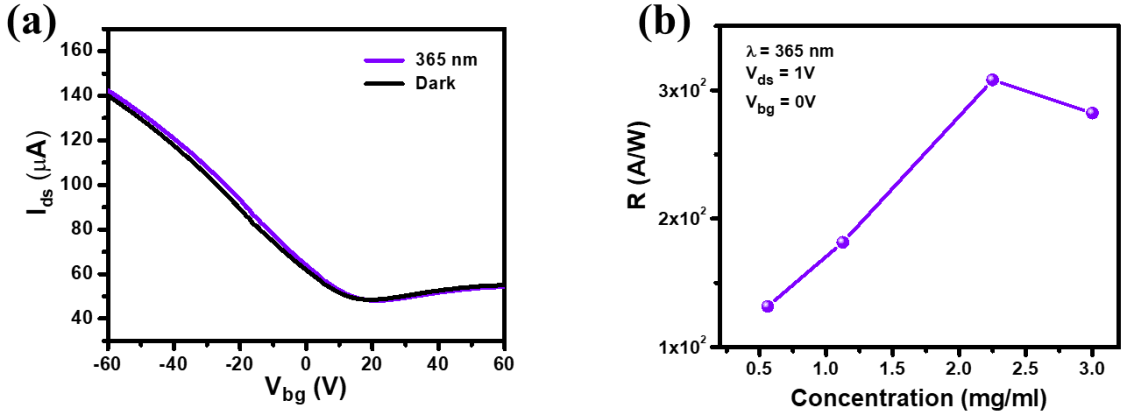

**Figure S 3(a)** Transfer characteristics of the hybrid device in dark and in illumination. $V_{ds} = 0.1V$, $\lambda = 365$ nm, $P_{LED} \sim 3.21$ μW/cm$^2$ **(b)** Photoresponsivity (R) as a function of concentration of WS$_2$ QDs at $V_{ds} = 1V$, $V_{bg} = 0V$, $\lambda = 365$ nm, $P_{LED} = 0.8$ μW/cm$^2$.

## 5. Responsivity, Noise equivalent power (NEP) and the specific detectivity (D*)

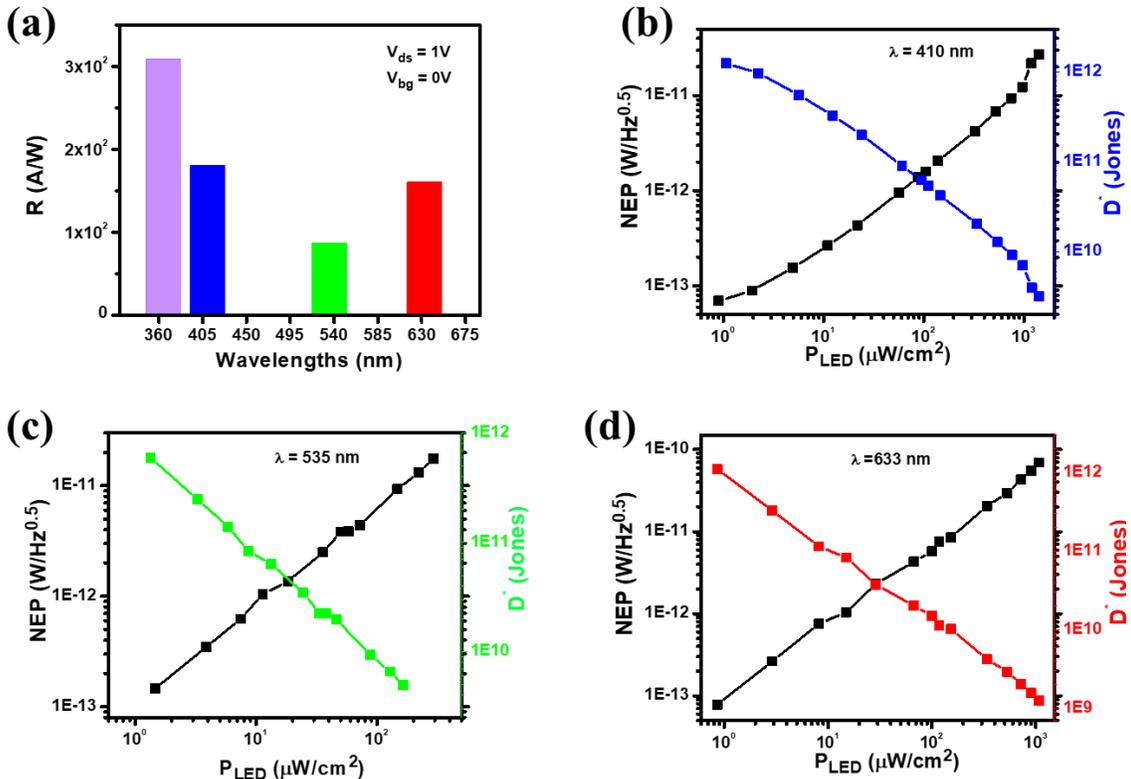

**Figure S 4(a)** Photoresponsivity (R) of the graphene-WS$_2$ QDs device at different wavelengths **(b), (c), (d)** NEP and specific detectivity (D*) of the hybrid device as a function of optical illuminations at $\lambda = 410$ nm, 535 nm, 633 nm.

## 6. **Density of states (DOS) calculation of WS$_2$ quantum dots (QDs)**

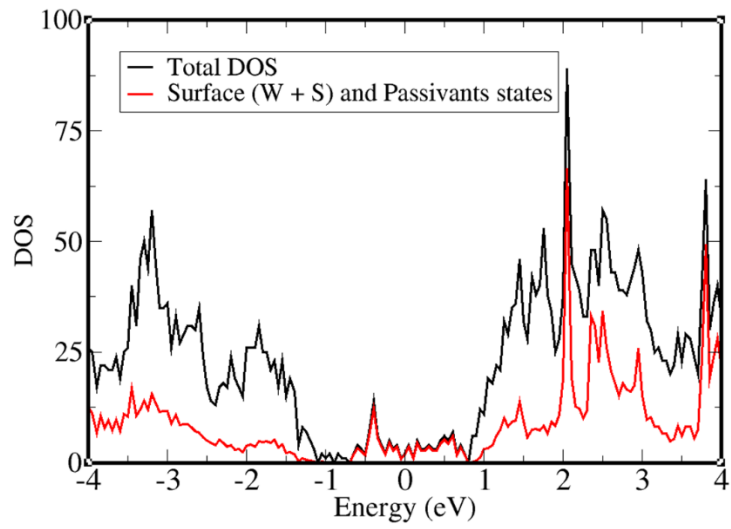

**Figure S 5**: Density of states as a function of energy for WS$_2$ QDs with lateral dimension 4.73 nm.